\documentclass[aps,prb,twocolumn,groupedaddress,superscriptaddress,amsfonts,amssymb,amsmath,floatfix,citeautoscript]{revtex4-1} 
\usepackage{graphicx} 
\usepackage[centering,hmargin=26mm,tmargin=26mm,bmargin=26mm]{geometry}
\usepackage{amsmath}
\usepackage{hyperref} 
\usepackage{multirow}
\usepackage{newtxtext}
\usepackage{comment}
\usepackage{booktabs}
\usepackage{xcolor} 
\usepackage{soul}
\usepackage{hyperref}
\usepackage{float}
\usepackage[normalem]{ulem}
\usepackage[textwidth=1.5cm,textsize=footnotesize]{todonotes}
\hypersetup{colorlinks, 
    linkcolor={blue!75!black!80!yellow},
    citecolor={blue!75!black!80!yellow}, 
    urlcolor={blue!75!black!80!yellow}
    }
\usepackage[cmintegrals]{newtxmath}

\makeatletter
\renewcommand\@make@capt@title[2]{%
    \@ifx@empty\float@link{\@firstofone}{\expandafter\href\expandafter{\float@link}}%
    \sffamily{\textbf{#1}}\@caption@fignum@sep#2
}%

\makeatother

\newcommand{\HarvardSEAS}{John A. Paulson School of Engineering and Applied Sciences, Harvard University, Cambridge, MA 02138, USA}

\def\NAS{NaSn$_2$As$_2$}
\def\PdO{PdCoO$_2$}
\def\PtO{PtCoO$_2$}
\def\taueph{$\tau_{\mathrm{eph}}$}
\def\tauee{$\tau_{\mathrm{ee}}$}

\def\EF{$E_F$}
\def\imee{Im$\Sigma_\mathrm{ee}$}
\def\imeph{Im$\Sigma_\mathrm{eph}$}
\def\tauMR{$\tau_{\mathrm{eph}}^\mathrm{MR}$}

\def\tautensor{$[\tau_{\mathrm{eph}}]_{ij}$}

\begin{document} 

\author{Yaxian Wang}\email{yaxianwang@seas.harvard.edu}\affiliation{\HarvardSEAS}
\author{Prineha Narang}\email{prineha@seas.harvard.edu}\affiliation{\HarvardSEAS}

\title{Anisotropic Scattering in Goniopolar Metal \NAS}

\date{\today}

\begin{abstract}
\noindent Recent experimental discoveries in axis-dependent conduction polarity, or \emph{goniopolarity}, have observed that the charge carriers can conduct like either electrons or holes depending on the crystallographic direction they travel along in layered compounds such as \NAS. To elucidate this unusual transport behavior, we present an \textit{ab initio} study of electron scattering in such systems. We study different microscopic scattering mechanisms in {\NAS} and present the electron-phonon scattering time distribution on its Fermi surface in momentum space, the anisotropy of which is proposed to be the origin of the axis-dependent conduction polarity. 
Further, we obtain the overall anisotropic lifetime tensors in real space at different electron chemical potentials and temperatures and discuss how they contribute to the macroscopic thermopower. While we find that the contribution of the in-plane and cross plane lifetimes exhibits a similar trend, the concave portion of the Fermi surface alters the electron motion significantly in the presence of a magnetic field, thus flipping the conduction polarity as measured via the Hall effect. Our calculations and analysis of {\NAS} also suggests the strong possibility of hydrodynamic electron flow in the system. Finally, our work has implications for anisotropic electron lifetimes in a broad class of goniopolar materials and provides key, general insights into electron scattering on open Fermi surfaces.

\end{abstract}

\maketitle
Transport in metals and semimetals is extremely sensitive to the dynamics at (and near) the Fermi level. Therefore, the geometric shape, or topology, of the Fermi surface can determine the conductivity tensor following Mott's relation: $\sigma_{ij}(E_F)=q^2N(E_F)\tau_{ij}(E_F)m^{*-1}_{ij}(E_F)$, where $q$ is the electron charge, $N$ the carrier density, $\tau_{ij}$ and $m^{*-1}_{ij}$ the scattering time and reverse effective mass tensors evaluated at the Fermi level {\EF}~\cite{cutler1969observation}. The latter can be visualized as the curvature of the energy band or the Fermi surface~\cite{jan1968effective}, whose topology can dramatically alter the electron transport, and thus give rise to exotic phenomena~\cite{kumar2019extremely,arnold2016negative,he2019nonlinear,besser2019pressure,cook2019electron,ali2016butterfly,pavlosiuk2017fermi,singha2018probing}. 

Recent findings in axis-dependent conduction polarity, or \textit{goniopolarity}, reveal that charge carriers can be `entangled' with the direction they travel in layered compound {\NAS},~\cite{he2019fermi} originating from its hyperboloid open Fermi surface. In \NAS, the effective mass has opposite signs along the in-plane \textit{x,y} and the cross plane $z$ axes. Delafossites {\PdO} and {\PtO}~\cite{ong2010unusual,eyert2008metallic} are expected to, based on their Seebeck coefficients, exhibit similar behavior due to their anisotropic effective mass tensors on their open, nearly cylindrical Fermi surfaces. While previous theoretical approaches successfully predict axis-dependent conduction polarity in various layered compounds~\cite{wang2020chemical}, the scattering time tensor $\tau_{ij}$ has not been explored. Indeed, the full picture of transport is complex, even in the single band goniopolar metals. 

As examples of the rich transport physics in these materials,  the conduction polarities in {\NAS} measured by the Seebeck and Hall effect are opposite, and hydrodynamic electronic transport is observed in {\PdO}.~\cite{nandi2018unconventional,moll2016evidence,bachmann2019super} Further, {\NAS} was discovered to be superconducting by $s$-wave pairing,~\cite{goto2017snas,Ishihara2018supercond,yuwen2019enhanced,Cheng2018nodeless} and an optical study of {\PdO} reveals unexpected features due to the coupling of in-plane charge carriers with cross plane optical phonons~\cite{Homes2019perfect}. Moreover, these systems show a quasi-two-dimensional crystal structure and band topology, in which hydrodynamic electron flow has been discovered recently~\cite{cook2019electron,sulpizio2019visualizing,Ella2019,Holder2019}. In each of these cases, electron scattering with both electrons and phonons (especially the latter) leads to unconventional transport behavior, and significant deviation from the traditional Fermi liquid regime under certain conditions. These results naturally raise the question as to whether assuming the relaxation time to be a constant scalar in goniopolar materials is valid. While the cross section of Fermi surface in the $xy$ plane of {\PdO} is nearly hexagonal, the star shaped Fermi surface in {\NAS} has both convex and concave segments, which as we will show later lead to even more complex unexpected transport behaviors. 

\begin{figure*}
    \centering
    \includegraphics[width=0.85\linewidth]{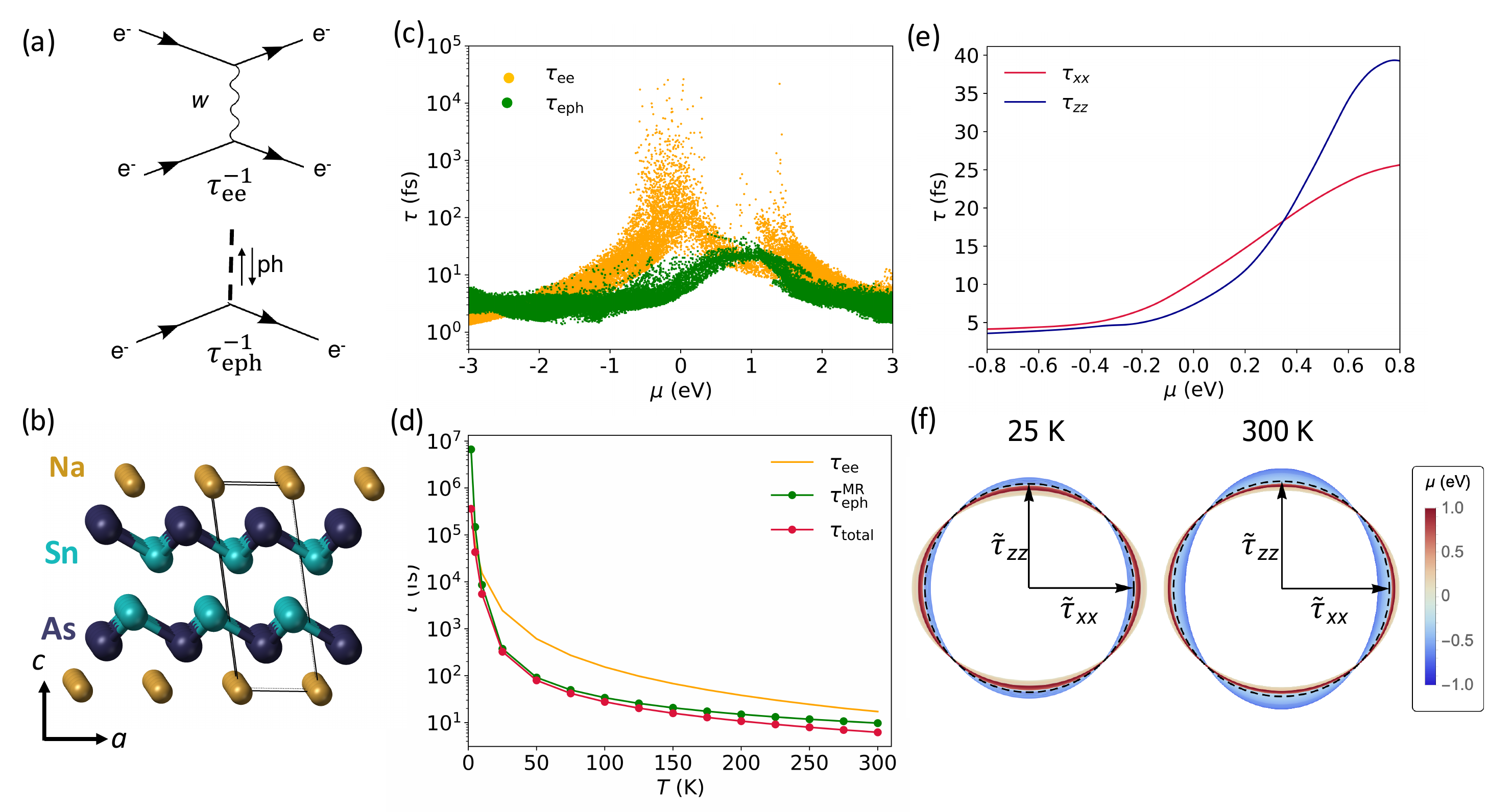}
    
    \caption[width=\textwidth]{(a) Feynman diagrams showing the vertices of electron-electron scatting through the screened Coulomb interaction and via a phonon. (b) Lattice structure of the primitive cell of {\NAS} viewed from [1 0 0], highlighting the vdW layers and Na atoms as electron donors. (c) Comparison of e-ph and e-e scattering times with dependence of electron chemical potential $\mu=\varepsilon-E_F$, highlighting that e-ph interaction is dominating near the {\EF}. The wide range of $\tau$ distribution is because electronic states at wavevectors across the Brillouin zone have different scattering rates. (d) Temperature dependence of {\tauMR}, {\tauee}, and $\tau_{\mathrm{total}}$ obtained by Matthiessen's rule. (e) Axis-dependent momentum relaxing {\tautensor} with dependence of electron chemical potential calculated at 300~K, showing that both $\tau'_{xx}=\frac{d\tau_{xx}}{dE}$ and $\tau'_{zz}=\frac{d\tau_{zz}}{dE}$ have positive signs at {\EF} with their positive slopes. (f) Normalized $\tilde{\tau}_{xx}$ and $\tilde{\tau}_{zz}$ at various electron chemical potentials calculated at 25~K and 300~K, showing the scattering time anisotropy is more significant at high temperatures. Black dashed circle is the isotropic reference as a guide.}
     \label{fig:tau_plot}
\end{figure*}

To elucidate the electron scattering physics of such open Fermi surface systems, in this \emph{Letter}, we study {\NAS} as a `typical' goniopolar metal and establish from first principles the microscopic origins of electron scattering. Furthermore, for the first time, we establish the anisotropic lifetime tensor $\tau_{ij}$ to investigate the axis-dependent scattering times, i.e. $\tau_{xx}$ and $\tau_{zz}$. We present the electron-electron scattering time ({\tauee}) and momentum relaxing electron-phonon scattering time ({\tauMR}) with dependence on both temperature and electron chemical potential, showing that the electron-phonon coupling dominates the scattering near the Fermi energy. We show how the anisotropic electron lifetimes contribute to the conduction polarity and modify the macroscopic thermopower reported in the earlier work~\cite{he2019fermi} on this material. To address the opposite sign of the Hall coefficients, we calculate the electron velocity evolution under a magnetic field, and find that the real space electron orbits self-intersect due to the concave segments of the Fermi surface~\cite{MRfermi,ong1991geometric}. This work sheds light on the anisotropic electron lifetimes in goniopolar materials, and provides important first principles insights into electron scattering microscopics on open Fermi surfaces.

\textit{\textbf {Scattering mechanisms}}. $-$
In a metal with a considerable electron density of states at the Fermi level, there are two primary scattering channels, electron-electron (e-e) scattering and electron-phonon (e-ph) scattering. We evaluate both scattering matrices from first principles calculations following a similar approach to that presented in earlier work.~\cite{su2019ultrafast,narang2017effects,brown2016nonradiative,garcia2020optoelectronic,coulter2019uncovering,coulter2018microscopic,ciccarino2018dynamics,narang2016cubic} Fig.~\ref{fig:tau_plot}(a) shows the vertices of 
e-e scattering through the screened Coulomb interaction (\textit{w}) considered here, as well as e-ph scattering involving an electron and a phonon (ph). The electron lifetimes for such events are determined by the imaginary part of electron self energy as $\tau_{\mathrm{ee}}=\hbar/2\mathrm{Im}\Sigma_{\mathrm{ee}}$ and $\tau_{\mathrm{eph}}=\hbar/2\mathrm{Im}\Sigma_{\mathrm{eph}}$, and $\tau_{\mathrm{total}}$ can be obtained by Matthiessen's rule (\emph{Methods} in Supplemental Material). While {\tauee} can be obtained from the electron quasiparticle self energy, {\taueph} relies on the details of crystal momentum, i.e. on the phonon dispersion. {\NAS} ($R\bar{3}m$, space group No.166) has a van der Waals (vdW) Zintl phase (Fig.~\ref{fig:tau_plot}(b)), with 5 atoms in a rhombohedral primitive cell.~\cite{arguilla2016nasn2as2} There is  one electron band crossing {\EF} in {\NAS} (Fig.~S1(b) in Supplemental Material) and the system has 15 phonon branches (Fig.~S1(a) Supplemental Material) which can be divided into three groups: 3 acoustic modes, 3 lower energy optical modes and 9 higher energy optical ones. There is no energy separation between the acoustic and optical modes, i.e. there is no \textit{ao} gap, but a big energy gap of about 10~meV exists between the lower and higher energy optical modes (Fig.~S1(a) in Supplemental Material).

A brief comparison between {\imee} and {\imeph} shows that e-e scattering is prominent along the cross plane ($\Gamma-Z$) direction. This is because {\NAS} has a quasi-two-dimensional structure with the buckled Sn-As honeycomb layers bonded by the van der Waals (vdW) force, which leads to optical phonons with little dispersion along $\Gamma-Z$. However, the partial filling of Sn $s/p_z$ orbitals gives rise to a highly dispersed electron band even along the stacking direction. A closer examination of this electron band along the high symmetry path in the Brillouin zone shows {\imeph} dominates most wavevectors along the in-plane direction ($\Gamma-$L and $\Gamma-$F) (Fig.~S1(c)). Given that the three acoustic phonon branches have comparable slopes along all directions, the enhanced {\imeph} along the in-plane direction indicates that the optical phonon modes also contribute to the electron-phonon scattering in addition to the acoustic ones.  
\begin{figure*}
    \centering
    \includegraphics[width=0.6\linewidth]{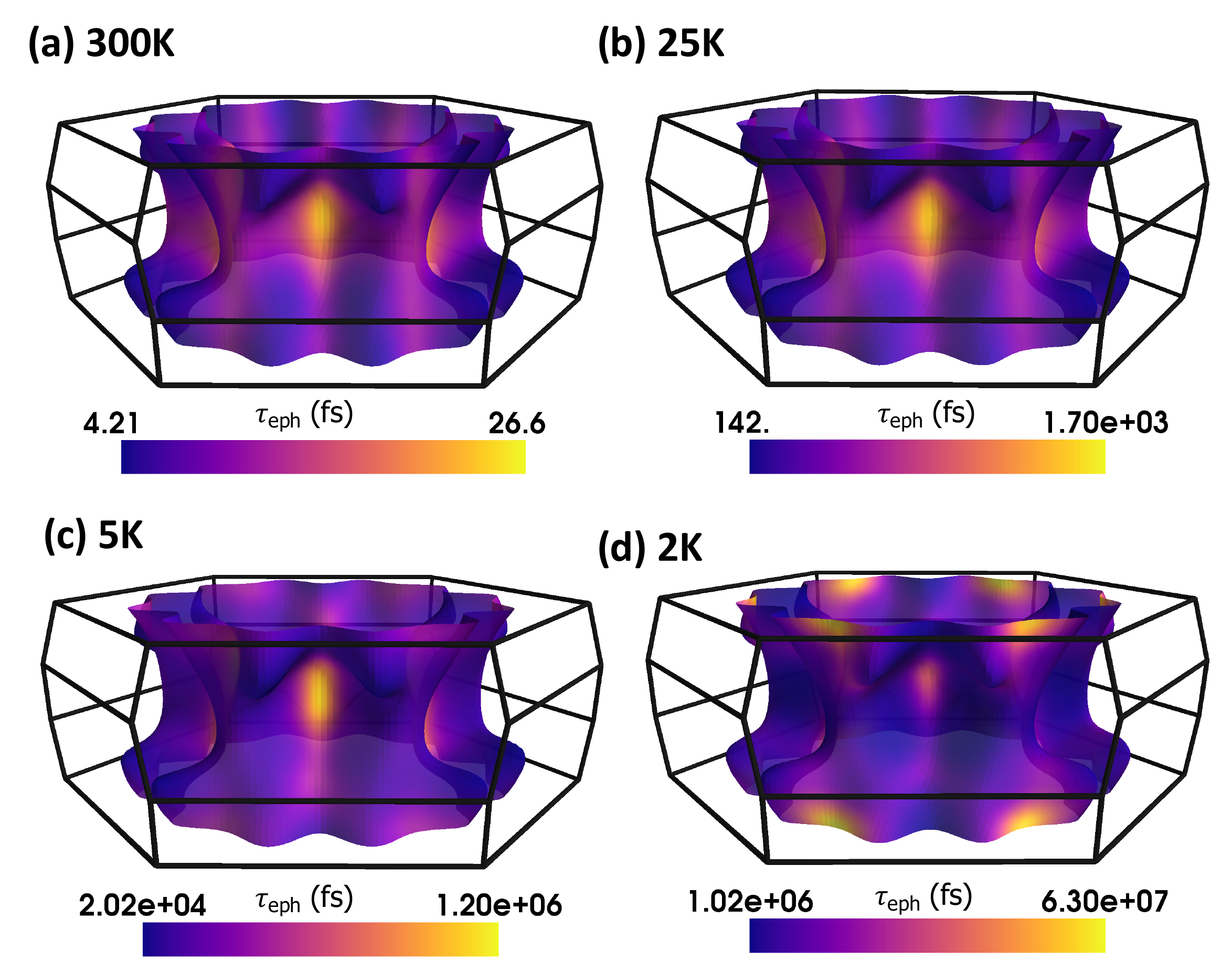}
    \caption[width=\textwidth]{Anisotropic {\tauMR} (fs) distribution on the Fermi surface, highlighting the appearance of the large lifetime regions across the Fermi surface with decreasing temperature (a) 300~K, (b) 25~K, (c) 5~K, and (d) 2~K. Similar to what has been observed on the open tube-shaped hole Fermi surface in WP$_2$~\cite{coulter2018microscopic}, there is a significant disparity between the lifetimes at low temperatures near 25~K. 
}
    \label{fig:FS_total}
\end{figure*}

While a qualitative comparison can be made by inspecting the high symmetry path, the strong anisotropy of this material and the complexity of its Fermi surface require knowledge of contributions from all the electronic states. The electron-electron and electron-phonon scattering times at wavevectors sampled across the whole Brillouin zone and at energies above and below the Fermi level are shown in Fig.~\ref{fig:tau_plot}(c). {\tauee} peaks at {\EF}, and its rapid reduction away from {\EF} can be described by a $(\varepsilon-E_F)^2$ dependence, indicating a Fermi liquid feature similar to that of simple metals (See Fig.~S2 for comparison with Al and Cu). However, there is another peak located at around $E_F+1.6$~eV, which can be due to the second lowest energy conduction band that has a high electronic density of states (Fig.~S1(b) right panel). Moreover, unlike Al and Cu, where {\taueph} clearly dominates in a wide energy range, in {\NAS} there is an energy window at around $E_F+1.3$~eV where {\tauee} can be smaller than {\taueph} for electrons with certain momenta. This is also indicated by the large {\imee} along the cross plane direction where the partially filled $s/p_z$ states can result in strong e-e coupling.

Prior work has shown that {\NAS} has a negligible magnetoresistance~\cite{he2019fermi}, indicating that there is only one microscopic length scale near the Fermi surface. Therefore, in the present work, we focus on the anisotropic {\taueph} within $\pm$0.8~eV of \EF. To get the momentum relaxing {\tauMR} that is more relevant to the transport properties, we account for the small angle scattering which commonly happen at low temperatures by weighing the electron scattering rate $\tau^{-1}_{\rm{eph}}(n\textbf{k})$ with $v_{n\textbf{k}}\cdot v_{n\textbf{k}}$ (\textit{Methods} in Supplemental Material). The temperature dependence of {\tauee}, {\tauMR}, together with $\tau_{\mathrm{total}}$ obtained by Matthiessen's rule is shown in Fig.~\ref{fig:tau_plot}(d). Without altering the electron chemical potential, {\tauMR} determines the electron scattering time from 5~K to 300~K, especially in the medium temperature range $25-100$~K. Normally, the resistivity that is governed by the interactions between electrons and acoustic phonons is expected to have a $T^{1}$ dependence at temperatures above $\sim\Theta_D/5$, where $\Theta_D$ is the Debye temperature and has been evaluated to be $\sim198$~K in {\NAS} from specific heat measurement~\cite{he2019fermi}. In the temperature region from 50 to 300~K, we fit the data with $\tau^{-1}(T)=A+BT^\alpha$ and obtained $\alpha\approx1.18$, showing a substantial deviation from the $T^{1}$ dependence. This further demonstrates that the optical phonon modes also play an important role in scattering electrons in this system. 

At very low $T (<<\Theta_D$), the resistivity of a metal usually follows $T^{5}$, in which $T^3$ is from the phonon density of states and $T^2$ is from the small angle scattering, where $\theta \approx  \vec{q}/ \vec{k} \approx T/\Theta_D$. Below 50~K ($2-25$~K in this context), we fit the data with similar $\tau^{-1}(T)=A'+B'T^{\alpha'}$ and obtained $\alpha'\approx3.89$. 
Possible reasons that it deviates from the $T^{5}$ dependence here include the following. First, in the present work, the momentum relaxing electron-phonon scattering time {\tauMR} is weighed by $1-\mathrm{cos}\theta$ when integrating $\tau^{-1}_\mathrm{eph}(n\textbf{k})$ at a certain temperature and thus could have less significant temperature dependence. Second, {\NAS} is a quasi-two-dimensional material, whose phonon density of states is expected to follow a temperature dependence between $T^2$ and $T^3$, the latter of which is typical of a three-dimensional solid. 


The macroscopic transport properties are determined mostly by the states near {\EF}; therefore we present the anisotropic electron-phonon lifetimes in momentum space across the Fermi surface at different temperatures in Fig.~\ref{fig:FS_total}. The Fermi surface of {\NAS} has an open concave shape which is consistent with previous work,~\cite{he2019fermi} although there is no pocket at the Z point without $n-$type doping. Though the Fermi surface features long lifetimes at similar regions at elevated temperatures (5K and above), the strong contrast, i.e. the range of 3 orders of magnitude, of the lifetimes at a `moderate' temperature 25~K is noteworthy. Similar spots of long {\tauMR} have also been proposed for the hole pocket of WP$_2$, a type-II Weyl semimetal. The Fermi surface of its valence band has an open quasi-two-dimensional `tube' shape, where the long-lived electrons are located at the waist of the tube. The strong anisotropy of {\tauMR} on the hole Fermi surface in WP$_2$ could contribute to hydrodynamic charge flow and the violation of Wiedemann-Franz law.~\cite{coulter2018microscopic} Meanwhile, we recall that a similar system {\PdO} has also been shown to exhibit hydrodynamic signatures, and it is also a single band goniopolar metal with a hexagonal open Fermi surface shape close to that of {\NAS}. With both of {\NAS} and {\PdO} having a quasi-two-dimensional crystal structure and band topology, we propose that {\NAS} could feature even richer transport behavior at low temperatures beyond axis-dependent conduction polarity~\cite{cook2019electron,sulpizio2019visualizing,Ella2019,Holder2019}. Our analysis of {\NAS} indicates a strong possibility of hydrodynamics in this goniopolar material and warrants further experimental investigation. Moreover, the distinct feature at 2~K of long lifetimes instead being located at the zone boundary rather than at the center is worth further study to help understand the superconducting phase in this vdW compound, which was verified at 1.3~K for stoichiometric {\NAS}~\cite{goto2017snas} and 2.1~K for Na$_{1+x}$Sn$_{2-x}$As$_2$~\cite{yuwen2019enhanced}.

\textit{\textbf{Axis-dependent $\mathbf{\tau_{eph}}$}}. $-$  The physical origin of the axis-dependent conduction polarity in {\NAS} is proposed to be the concave Fermi surface topology, which manifests as the opposite signs of its in-plane ($\alpha_{xx}$)  and cross plane ($\alpha_{zz}$) thermopower, each evaluated with the ratio of $\sigma'_{ij}$ ($\frac{d\sigma_{ij}}{dE}$) over $\sigma_{ij}$ at $\varepsilon=E_F$ following:
\begin{equation}
     \alpha_{ij}=-\frac{\pi^2k_B^2T}{3q}\times \left\{\left.\frac{N'}{N}\right\vert_{\varepsilon=E_F}+\left.\frac{\tau_{ij}'}{\tau_{ij}}\right\vert_{\varepsilon=E_F}+\left.\frac{[m_{ij}^{*-1}]'}{[m_{ij}^{*-1}]}\right\vert_{\varepsilon=E_F} \right\},
     \label{thermopower}
\end{equation}

where $k_B$ is the Boltzmann constant, and primed quantities denote derivatives with respect to energy. Here, the carrier density $N$ is a scalar, but both the inverse effective mass $m_{ij}^{*-1}$ and $\tau_{ij}$ are tensors. After establishing that {\tauMR} dominates the electron lifetimes in {\NAS}, we now examine the axis-dependent $[\tau_{\mathrm{eph}}]_{ij}$ and discuss its contribution to the overall thermopower. We obtain the anisotropic {\tautensor} by weighing each momentum relaxing scattering rate at different states by $v_{n\textbf{k}}\otimes v_{n\textbf{k}}$ (\textit{Methods} in Supplemental Material). Both $\tau_{xx}$ and $\tau_{zz}$ with dependence of electron chemical potential at 300~K are shown in Fig.~\ref{fig:tau_plot}(e). As mentioned earlier, we restrict the energy window to within 0.8~eV with respect to \EF, where we can safely assume that {\tauMR} determines the electron lifetimes. Both $\tau_{xx}$ and $\tau_{zz}$ have a positive slope near {\EF}, with an anisotropy ratio $\tau_{zz}/\tau_{xx}<2$. Examination of $\tau_{xx}'/\tau_{xx}$ and $\tau_{zz}'/\tau_{zz}$ confirms they have the same positive sign for $\varepsilon-E_F<0.8$~eV. Fig.~\ref{fig:tau_plot}(f) shows the comparison of principal components of {\tautensor} normalized as $\tilde{\tau}_{zz,xx}=\sqrt{\tau_{zz,xx}/(\sqrt{\tau_{zz}}\cdot \sqrt{\tau_{xx}})}$ at different temperatures. The change of the scattering time anisotropy versus electron chemical potential is much smaller at lower temperature, which suggests that the axis-dependent conduction polarity in {\NAS} does indeed originate from its Fermi surface topology instead of the electron scattering. However, the anisotropic lifetimes do contribute a substantial portion to the overall thermopower. According to previous work~\cite{he2019fermi}, only considering the band topology, the predicted $\alpha_{xx}$ is -2~$\mu$V/K at 300~K, which was measured to be -2 and -6~$\mu$V/K in two different samples. On the other hand, $\alpha_{zz}$ was predicted to be +21~$\mu$V/K while measured to be +8 and +10~$\mu$V/K~\cite{he2019fermi}. According to Eq.~\ref{thermopower}, the lifetime tensor will contribute $-\frac{\pi^2k_B^2T}{3e}\times\frac{\tau'_{xx}}{\tau_{xx}}$ to $\alpha_{xx}$ and $-\frac{\pi^2k_B^2T}{3e}\times\frac{\tau'_{zz}}{\tau_{zz}}$ to $\alpha_{zz}$, both of which are negative. Therefore including the contribution from the anisotropic scattering time tensor decreases the absolute value of $\alpha_{zz}$ but increases the absolute value of $\alpha_{xx}$, thus bringing the theoretical results in considerably better agreement with the experimental measurements. 

\begin{figure}
    \centering
    \includegraphics[width=\linewidth]{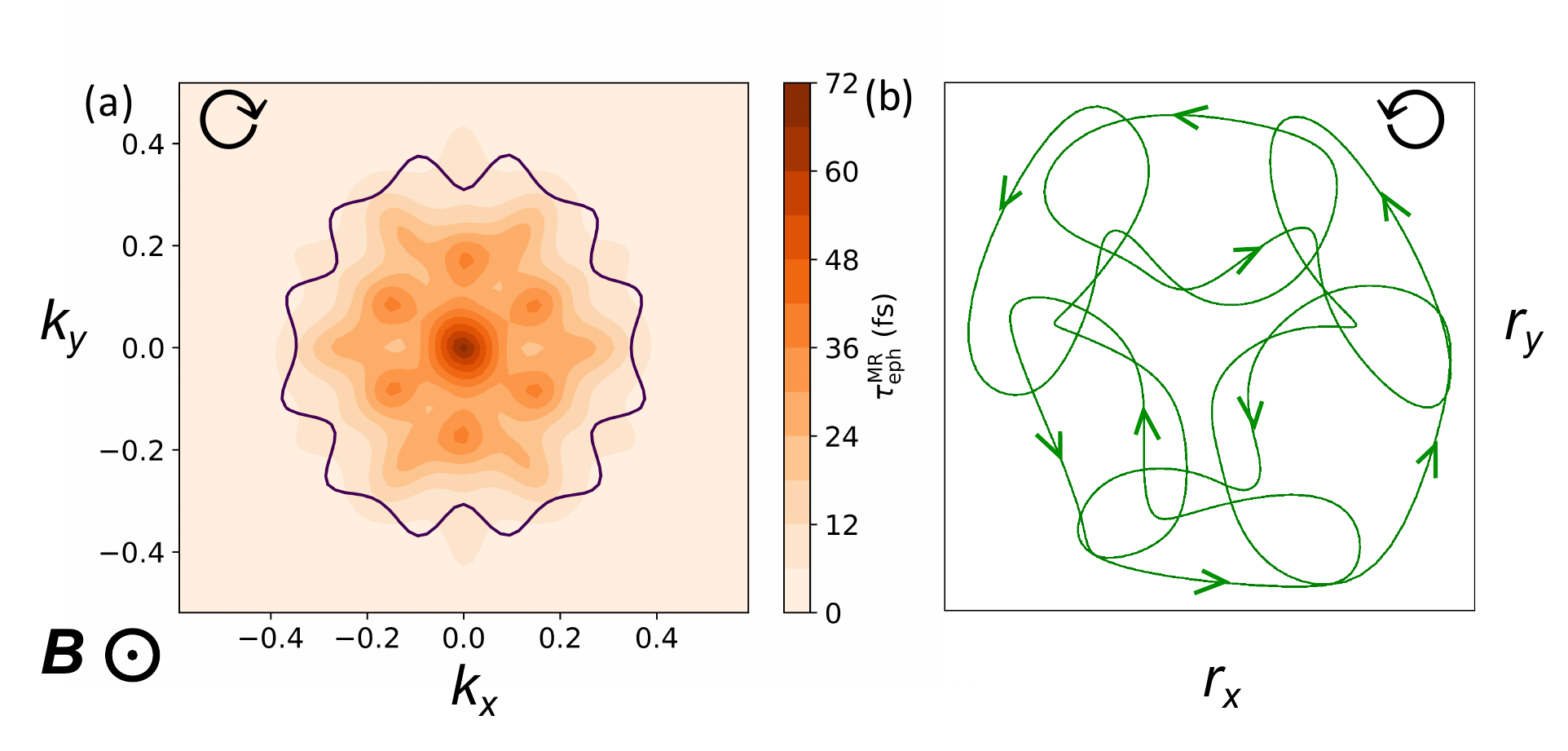}
    \caption[width=\textwidth]{(a) Fermi surface cross section at $k_z=0.25c^*$ plane, where $c^*$ is the cross plane reciprocal lattice vector, outlined by solid line in purple, with color map showing the scattering time distribution. (b) Electron orbits in real space for the Fermi surface shape in (a), highlighting the self-intersection due to the presence of concave areas on the Fermi surface loop. 
}
    \label{fig:Hall}
\end{figure}

\textit{\textbf{Hall conductivity.} $-$} While we conclude that the goniopolar thermoelectric response in {\NAS} originates from the anisotropic effective mass tensor instead of the scattering time tensor,  another question left unanswered is why the conduction polarity is flipped in the Hall effect measurement. Specifically, with negative $\alpha_{xx}$ indicating electron conduction along the in plane direction, the Hall coefficient is measured to be positive, indicating positive charge carriers, i.e. holes. To solve this puzzle, we explicitly calculate the electron velocity evolution under an applied magnetic field. 

Fig.~\ref{fig:Hall}(a) shows the Fermi surface as a loop on a slice of the $xy$ plane at mid point of $\Gamma-Z$, i.e. $k_z=0.25c^*$ where $c^*$ is the cross plane reciprocal lattice vector. Since we are moving away from the zone center, the Fermi surface shows much more complexity than a simple hexagonal shape. There are multiple concave and convex regions along $k_x$ and $k_y$. With color showing the distribution of {\tauMR} on such a slice, the long lifetime spots on the Fermi surface loop can be seen, which are the six concave segments with smaller curvature, as visualized by the color highlighting the same position in Fig.~\ref{fig:FS_total}. Applying a magnetic field along $k_z$ will impose a Lorentz force on the electron velocities, the evolution of which can be evaluated assuming a constant relaxation time (\textit{Methods} in Supplemental Material) and is shown in Fig.~\ref{fig:Hall}(b). The electron orbit in real space is clearly complicated and is subject to the subtlety of the Fermi surface. The most pronounced feature is that it self-intersects multiple times where the Fermi surface contains concave segments at several points. With arrows guiding the eye, we can see that the circulation of electrons has opposite directions. According to Ong~\cite{ong1991geometric}, the Hall conductivity is determined by the circulation area of the `Fermi surface' in the mean free path space. Due to the lattice symmetry, we treat the in-plane electron scattering time as isotropic~\cite{varnavides2020generalized}. Thus the evolution of \textbf{v}(\textbf{k}) can be directly mapped into the mean free path space since $\textbf{r}(\textbf{k})=\textbf{v}(\textbf{k})\tau$. Here, the outer and inner loops show opposite circulation, i.e. clockwise in the momentum space but anticlockwise in the mean free path space, resulting in the sign flipping of the Hall conductivity.

\textit{\textbf{Conclusions and Outlook.}$-$} In summary, we present a first principles study of electron scattering on an open Fermi surface. We study a single band metal {\NAS}, which was shown to have opposite conduction polarity along in-plane and cross plane directions originating from its band topology, to reveal how the electron lifetimes affect its macroscopic transport properties. We evaluate the electron lifetimes due to scattering with both electrons and phonons with energy and momentum resolution and find that the electron-phonon scattering dominates the electron lifetimes within 0.8~eV of the Fermi level, in temperature range of 5~K to 300~K. We further study the anisotropic electron-phonon lifetimes $\tau_{xx}$ and $\tau_{zz}$ at different electron chemical potentials, and discuss their contribution to the macroscopic thermopower and the overall axis-dependent conduction polarity.

Finally, we address the opposite conduction polarity observed in the Hall effect by calculating the real space electron orbits in the presence of an external magnetic field, and show how the self intersection leads to opposite circulation which flips the sign of Hall coefficients. These together can serve as a foundation for understanding the electron scattering rates in materials with axis-dependent conduction polarity, as well as drawing a more general conclusion on exotic transport behaviors in systems with nontrivial Fermi surface topology. Interestingly, our analysis of {\NAS} suggests the strong possibility of hydrodynamics in this (and similar goniopolar) materials, that we expect will spark experimental investigation. \\

\noindent \textbf{Acknowledgements} We  acknowledge  fruitful   discussions with Christina A.C. Garcia and Georgios Varnavides (Harvard), Prof. Wolfgang Windl (Ohio State University), and Prof. Claudia Felser (Max Planck for Chemical Physics of Solids). This work is supported by the STC Center for Integrated Quantum Materials, NSF Grant No. DMR-1231319.
This research used resources of the National Energy Research Scientific Computing Center, a DOE Office of Science User Facility supported by
the Office of Science of the U.S. Department of Energy under Contract No. DE-AC02-05CH11231, resources at the Research Computing 
Group at Harvard University as well as the Ohio Supercomputer Center.
P.N. is a Moore Inventor Fellow and gratefully acknowledges support through Grant GBMF8048 from the Gordon and Betty Moore Foundation.

\bibliography{ref.bib}

\end{document}